\documentclass[1p]{elsarticle}

\usepackage{amsmath,amssymb,xspace,graphicx,upgreek}
\usepackage[utf8]{inputenc}
\usepackage{t1enc}

\renewcommand{\vec}[1]{\ensuremath{\boldsymbol{#1}}\xspace}

\renewcommand{\d}{\ensuremath{\mathrm{d}}\xspace}

\renewcommand{\O}{\ensuremath{\langle\mathcal{O}^2(L)\rangle}\xspace}
\newcommand{\OR}{\ensuremath{\langle\mathcal{O}^2(R)\rangle}\xspace}
\newcommand{\D}{\ensuremath{\mathcal{D}}\xspace}
\newcommand{\kB}{\ensuremath{k_{\text{B}}}\xspace}
\newcommand{\TDG}{\ensuremath{\tilde{T}_{\text{DG}}}\xspace}
\newcommand{\TR}{\ensuremath{\tilde{T}_{\text{R}}}\xspace}
\newcommand{\EJ}{\ensuremath{E_{\text{J}}}\xspace}

\newcommand{\xih}{\ensuremath{\xi_{\text{h}}}\xspace}

\newcommand{\ket}[1]{\left| #1\right\rangle}
\newcommand{\bra}[1]{\left\langle #1\right|}
\newcommand{\cs}{\ensuremath{c_\text{s}}\xspace}
\newcommand{\rhos}{\ensuremath{\rho_\text{s}}\xspace}

\DeclareMathOperator{\J}{J}

\begin{document}

\title{Non-local order in Mott insulators, Duality and Wilson Loops}
\author[tum]{Steffen Patrick Rath\corref{cor}}
\ead{steffen.rath@ph.tum.de} 
\author[tum]{Wolfgang Simeth}
\author[mpq]{Manuel Endres}
\author[tum]{Wilhelm Zwerger}
\ead{zwerger@ph.tum.de}
\cortext[cor]{Corresponding author, phone +49\,89\,289\,12365, 
fax +49\,89\,289\,12638}
\address[tum]{Technische Universit\"at M\"unchen, James-Franck-Stra{\ss}e 1, 85748 Garching, Germany}
\address[mpq]{Max-Planck-Institut f\"ur Quantenoptik, 85748 Garching, Germany}

\begin{abstract}
It is shown that the Mott insulating and superfluid phases of bosons in an
optical lattice may be distinguished by a non-local 'parity order parameter'
which is directly accessible via single site resolution imaging. In one
dimension, the lattice Bose model is dual to a classical interface roughening
problem. We use known exact results from the latter to prove that the parity
order parameter exhibits long range order in the Mott insulating phase,
consistent with recent experiments by Endres et al.~[Science \textbf{334}, 200
(2011)]. In two spatial dimensions, the parity order parameter can be expressed
in terms of an equal time Wilson loop of a non-trivial $U(1)$ gauge theory in
$2+1$ dimensions which exhibits a transition between a Coulomb and a confining
phase. The negative logarithm of the parity order parameter obeys a perimeter
law in the Mott insulator and is enhanced by a logarithmic factor in the
superfluid.
\end{abstract}
\maketitle

\section{Introduction}
\label{sec:intro}

A central assumption underlying both classical and quantum physics
is that its basic laws are {\it local}, i.e., that the fundamental equations
can be expressed in terms of relations between physical observables
at a given point in space and time~\cite{haag1996}. Symmetries and 
different types of order are thus related to the behavior 
of correlation functions of local observables. In particular, a qualitative 
change in macroscopic properties is typically associated 
with the appearance of long range order in some local observable 
$\hat{\mathcal{O}}(x)$. Its two-point correlation  
\begin{equation}
  \label{eq:LRO}
  \langle \hat{\mathcal{O}}(x) \hat{\mathcal{O}}(y)\rangle \to \mathcal{O}^2(\infty)\ne 0
\end{equation} 
thus approaches a finite constant as $\vert x-y\vert$ goes to
infinity.  In recent years, a lot of interest has focussed on systems where this
standard characterization of different phases of matter by finite values of some
local order parameter $\hat{\mathcal{O}}(x)$ fails. This is the case, e.g., in
the Quantum Hall Effect, a paradigmatic example of a topological insulator.  It
is an incompressible state described by a Chern-Simons theory which has gapless
excitations associated with chiral edge currents~\cite{wen2004,froh1993}. 

Our aim in the present paper is to study non-local orders that may be used to
characterize Mott insulating phases of lattice bosons.  Quite generally, Mott
insulators are defined by their incompressibility $\kappa=\partial
n/\partial\mu\equiv 0$, a response function that is not associated with long
range order in any local observable $\hat{\mathcal{O}}(x)$. Specifically, we
focus on Mott insulators which do not break any lattice symmetries due to, for
instance, the formation of a commensurate charge density wave~\cite{lee1990} or
due to Neel order in the spin degrees of freedom, as happens in Mott-Heisenberg
insulators~\cite{gebh1997} realized in undoped high-temperature
superconductors~\cite{lee2006}.  To investigate possible non-local orders that
may exist in otherwise featureless Mott insulators, the particular case of
bosons in an optical lattice is of special interest.  By a simple change of the
lattice depth, they may be tuned to undergo a superfluid (SF) to Mott insulator
(MI) transition~\cite{grei2002}. Moreover, thanks to the direct accessibility of
the atomic distribution by optical imaging methods which allow measuring in situ
density distributions~\cite{geme2009} and even arbitrary density correlations at
the single-atom level~\cite{sher2010,bakr2010}, they also provide a new
perspective on the microscopic details of the involved states.  The question of
whether some non-local order exists in the MI  phase of bosons in an optical
lattice has been addressed before by Berg et al.~\cite{berg2008} in the
particular case of one dimension. While their main focus has been the study of
hidden 'string-order' that appears in the presence of repulsive interactions of
finite range, they have shown via Bosonization that even in the much simpler
situation of pure on-site repulsion, there is a non-local observable which takes
finite values in the MI and is zero in the SF.  Using single site imaging, the
associated 'parity order parameter' (POP) has been measured by Endres et
al.~\cite{endr2011}. For lengths of up to eight lattice spacings their results
were consistent with the theoretical expectation of a non-vanishing parity order
in the MI and an algebraic decay to zero in the superfluid phase.

Our aim in the following is a detailed study of parity order in $d=1$ and also
in $d=2$ dimensions using duality transformations.  Specifically, in $d=1$, a
number of exact results for the parity order parameter are obtained from a
systematic expansion around the atomic limit and from analytical results for the
roughening transition of the equivalent classical interface model in two
dimensions~\cite{froh1981,forr1986,abra1986}. In the $d=2$ case, the lattice
bosons are dual to a three-dimensional $U(1)$ lattice gauge
theory. This mapping has first been discussed by
Peskin~\cite{pesk1978} starting from the classical three-dimensional XY model
and has been extended to the two-dimensional quantum XY model by Fisher and
Lee~\cite{fish1989a}. As will be shown here, the parity order parameter is
mapped in the dual model on a quantity which is 'more local by one dimension':
in $d=1$, it is mapped on a two-point correlation function for a local operator of the type in
Eq.~\eqref{eq:LRO} which shows an algebraic decay in the SF phase and converges
to a constant in the MI. In the $d=2$ case, the POP is mapped onto an equal time
Wilson loop in the dual gauge theory, which serves to distinguish the superfluid
and Mott insulating phases according to the dependence on the system size $L$.
In particular, we find a perimeter law dependence in the MI phase which leads
to an exponential decay of the POP while in the SF phase a logarithmic
correction to the perimeter law leads to a super-exponential decay.

\section{Bose-Hubbard Model and Parity Order}
\label{sec:model}

Ultracold bosons in an optical lattice provide a realization of the
Bose-Hubbard model, as suggested theoretically by Jaksch et al.~\cite{jaks1998}
and first realized experimentally by  Greiner et al.~\cite{grei2002}. The
associated many-body Hamiltonian~\cite{fish1989}
\begin{equation}
  \label{eq:HBH}
\hat{H}_{\text{BH}} = 
  -J  \sum_{\langle i,j\rangle} {\hat{a}^{\dagger}_{i} \hat{a}_{j} } 
  +\frac{U}{2}\sum_{i} {    \hat{n}_{i} \left( \hat{n}_{i} -1 \right) } 
  \ ,
\end{equation}
describes the competition between a kinetic energy which involves hopping to
nearest neighbor lattice sites with amplitude $J>0$ and an on-site repulsive
interaction $U>0$ which leads to an increase in energy if atoms hop to sites
which are already occupied. Specifically, the operator $\hat{a}_i$ destroys a
boson in a single particle Wannier state localized at lattice site $i$ and the
associated occupation number operator $\hat{n}_i=\hat{a}_i^\dagger\hat{a}_i$ has
eigenvalues $0,1,2,\dots$. We consider this model in both one and two spatial
dimensions, specializing to the case of a simple quadratic lattice in the latter
case.  In either dimension, the ground state of the Bose Hubbard model exhibits
a continuous transition between a superfluid and a Mott insulating state which
may be tuned either by changing the ratio $J/U$ at fixed density or by changing
density at fixed $J/U$ via the dimensionless chemical potential $\mu/U$
~\cite{fish1989}. The universality class of this quantum phase transition is
different in both cases. For the  density driven Mott transition, the
associated dynamical scaling exponent is $z=2$~\cite{sach1999}.  By contrast,
changing the ratio $J/U$ at fixed integer density $\bar{n}=1,2,\ldots$, the
transition at the tip of the Mott lobes has $z=1$, i.e., it is described by an
$O(2)$-model in $D=d+1$ dimensions which possesses  a formal relativistic
invariance~\cite{sach1999}. It is this type of transition which will be
considered in the following. 

The superfluid phase of the Bose Hubbard model is characterized by a
conventional, local observable  $\hat{\mathcal{O}}(x)=\hat{a}_i$ associated with
long range order in the one-particle density matrix, with
$\mathcal{O}^2(\infty)\equiv n_0$ as the condensate density\footnote{This would
not be true if the bosons were charged as, e.g., the Cooper pairs of a
conventional superconductor, where the role of $\hat{\mathcal{O}}(x)$ is
expected to be taken by the bi-Fermion operator
$\hat{\psi}_{\uparrow}\,\hat{\psi}_{\downarrow}\, (x)$. The associated
correlation function  $\langle \hat{\mathcal{O}}^{\dagger}(x)
\hat{\mathcal{O}}(y)\rangle$, however, is not gauge invariant and therefore does
not constitute a proper order parameter, as noted by Wen~\cite{wen2004}.}.  This
order can be observed in a direct manner in time-of-flight
images~\cite{grei2002}, which measure the momentum distribution as the Fourier
transform of the one particle density matrix~\cite{bloc2008}. The excellent
quantitative agreement between the measured absorption images after
time-of-flight and precise quantum Monte Carlo calculations which include both
the effects of finite temperature and the harmonic trap
potential~\cite{trot2010}, shows that the Bose-Hubbard
Hamiltonian~(\ref{eq:HBH}) provides a faithful description of cold atoms in an
optical lattice.  In the Mott insulating phase, the correlation function of  the
local bosonic field operator vanishes exponentially like
$\langle\hat{a}^{\dagger}(x)\hat{a}(0)\rangle\sim\exp(-\vert x\vert/\xi)$
with a correlation length $\xi\sim 1/\Delta$ that diverges like the inverse of
the Mott gap $\Delta$.  The Mott phase is characterized by its incompressibility
and has no associated order parameter, evolving in a continuous manner from a
thermally disordered state as the temperature is lowered below the Mott gap
$\Delta$. Note that the observation of peaks in the noise correlations
$\langle\hat{n}(\vec{x})\hat{n}(\vec{x'})\rangle$ in the Mott phase
after time--of--flight
 by F\"olling et al.~\cite{foll2005} does {\it not} reflect any
long range order: they are a consequence of the fact that the Fourier components
$\sum_{\vec{R}}n_{\vec{R}}\exp[i(\vec{k}-\vec{k'})\cdot\vec{R}]$ of the
\emph{average} density $n_{\vec{R}}=\langle\hat{n}_{\vec{R}}\rangle$ are of
order $N$ at wave vector differences $\vec{k}-\vec{k'}=m(\vec{x}-\vec{x'})/\hbar
t$ which are equal to a reciprocal lattice vector $\vec{G}$.

As will be discussed in the following,  the fundamental
difference between the superfluid and Mott insulating phase in terms of their
compressibility implies that there is a non-local observable which behaves in a
characteristically different manner in both phases.  By a straightforward
generalization of the parity order in 1d introduced by Berg et
al.~\cite{berg2008}, the non-local order is defined as
\begin{equation}
  \langle\mathcal{O}^2(\mathcal{D})\rangle
  =\left\langle
  e^{i\pi\sum_{i\in\mathcal{D}}(\hat{n}_i-\bar{n})}
  \right\rangle
  =\left\langle\prod_{i\in\mathcal{D}}
  (-1)^{\bar{n}}\hat{p}_i
  \right\rangle\ ,
  \label{eq:OP}
\end{equation}
where $\mathcal{D}$ is a spatial domain, i.e., an interval in $d=1$ and an area
in $d=2$, and $\hat{p}_i=(-1)^{\hat{n}_i}$ is the parity operator on lattice
site $i$. There are two important properties of this observable which should
be noted right away:  First of all, the observable is easily accessible in
experiments since, due to light-induced collision losses, quantum gas microscopes
directly measure the on-site parity rather than the actual occupation
numbers~\cite{bakr2010,sher2010}. As a second point, the observable must be
calculated and measured in an {\it open} domain $\mathcal{D}$ which is part of a
larger system, otherwise $ \langle\mathcal{O}^2(\mathcal{D})\rangle\equiv 1$
would trivially be equal to one due to conservation of particle number.   To
study the dependence on the size, we shall characterize the domain
\D by its linear extension $L$ measured in units of the lattice spacing.  In
analogy to the standard definition \eqref{eq:LRO} of long range order, the
parity order parameter (POP) \O exhibits long range order if
$\langle\mathcal{O}^2(\infty)\rangle$ is finite.

This kind of order parameter is analogous to those studied in Ising models with
a local gauge invariance, which have no conventional phase transitions to states
with long range order, yet may exhibit different phases distinguished by
non-local order parameters ~\cite{wegn1971,frad1978,tasa1991} (see also
section~\ref{sec:outlook}). In the context of cold atoms, a more complicated
'string order' parameter was introduced, which characterizes a Haldane insulator
that can form in one dimensional systems with longer range
interactions~\cite{torr2006}. Our focus is on the behavior of the parity order
parameter at the conventional SF--MI transition, not only in
1d~\cite{berg2008,endr2011} but also in 2d.  In particular, we will use a
duality transformation to show that in two dimensions parity order is related to
an equal time Wilson loop in a non-trivial $U(1)$ gauge theory which exhibits
different behavior as a function of system size $L$ in the MI and SF phases.

\section{Number fluctuations and area law}
\label{sec:quali}

\begin{figure}[htbp]
  \centering
  \includegraphics[width=.5\linewidth]{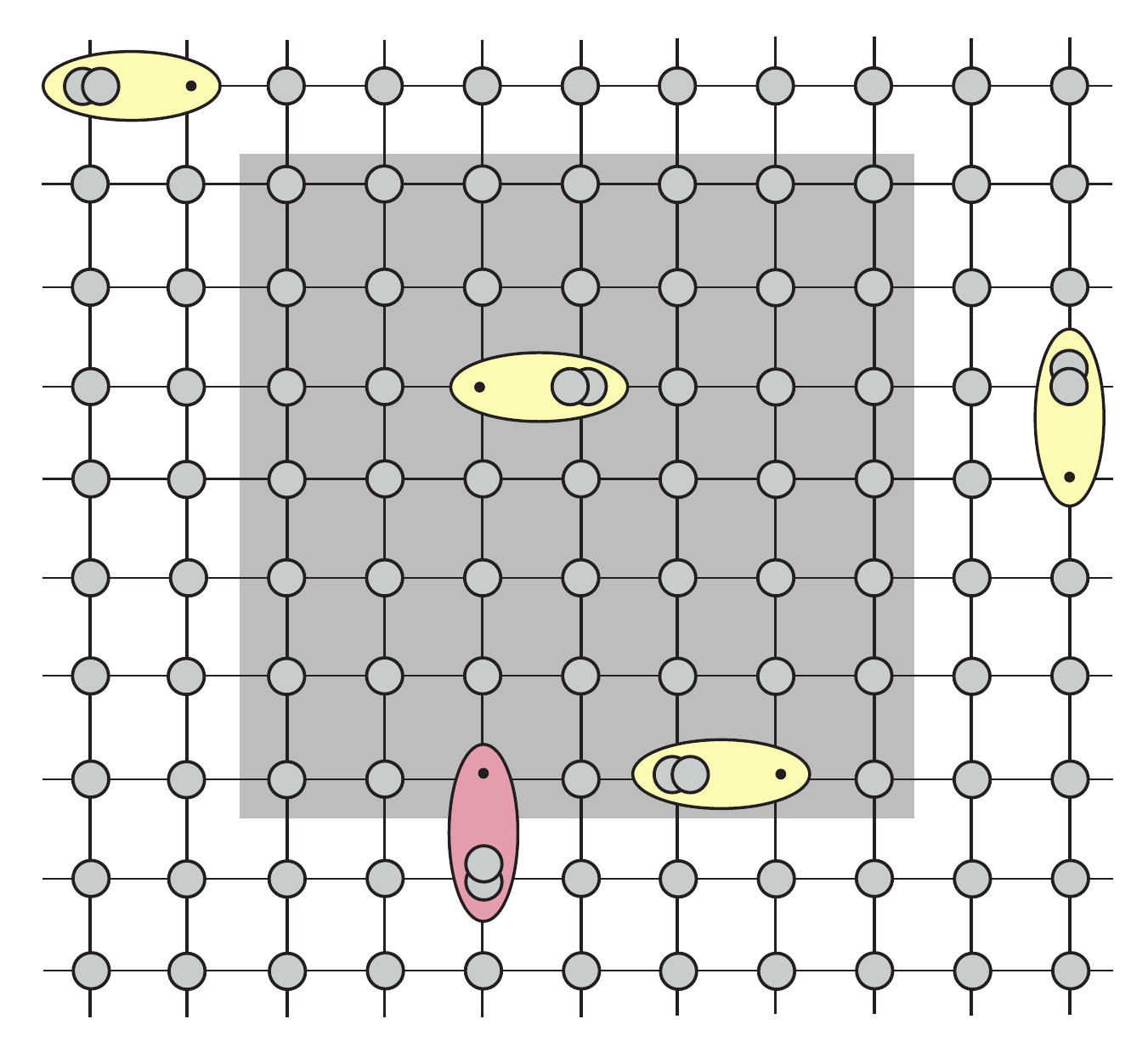}
  \caption{(Color online) Illustration of how the order parameter is reduced
  from unity for $d=2$. The grid lines indicate the lattice, bosons are
  represented by small circles. The domain $\D$, here taken to be a square, is
  shaded in gray.  The minus signs from pairs which are completely inside or
  outside the domain (yellow ellipses) cancel out so that there is no
  contribution while pairs which are separated by the domain boundary (red
  ellipse) contribute a minus sign which leads to a reduction of \O.}
  \label{fig:oreduction}
\end{figure}
To obtain a qualitative understanding of the dependence of the non-local order
defined in~(\ref{eq:OP}) on the size $L$ of the domain we start by giving some
qualitative arguments for the expected scaling behavior deep in the MI, where
$J/U\ll 1$. Starting with the atomic limit, it is obvious that $\O|_{J=0}=1$ in
any dimension since particle fluctuations are then completely frozen. For small
but finite $J/U$ and for the simple case of a MI with average density
$\bar{n}=1$, particle number fluctuations appear as pairs of empty and doubly
occupied sites. Now, for an arbitrary number of pairs which are completely
inside the domain $\mathcal{D}$, the associated two minus signs in the product
in Eq.~(\ref{eq:OP}) cancel. Only those pairs which are separated by the domain
boundary lead to a reduction of \O~\cite{endr2011}.  Intuitively one thus
expects the parity order $\O\sim\exp(-L^{d-1})$ to scale exponentially with the
area of the domain's boundary.  This intuitive expectation is supported by a
systematic perturbative calculation in an expansion around the atomic limit
$J/U=0$ which---as will be shown in section~\ref{sec:pert} below---yields
\begin{equation}
  \O = 1-8\bar{n}(\bar{n}+1)d L^{d-1}\left( \frac{J}{U} \right)^2+\cdots
  \label{eq:Oleadingorder}
\end{equation}
up to second order in $J/U\ll 1$.  Clearly the expansion is well defined only in
$d=1$, while in higher dimensions the effective expansion parameter $(J/U)^2\,
L^{d-1}$ is small only up to system sizes of order $(U/J)^{2/(d-1)}$.  

Further insight into the origin of the perimeter law for the decay of the parity
order can be gained by assuming that the expectation value in~(\ref{eq:OP}) may
be calculated within a Gaussian approximation such that 
\begin{equation}
  \O\approx e^{\langle(i\pi\sum_{i\in\D}\delta\hat{n}_i)^2\rangle/2}
  =e^{-\pi^2\langle \delta \hat{N}^2\rangle/2}\ , \label{eq:deltaN}
\end{equation} 
where $\delta \hat{n}_i=\hat{n}_i-\bar{n}$. Within this
approximation, $-\ln\O$ is simply a measure of the total number fluctuations
$\langle\delta \hat{N}^2\rangle$ in a  domain of size $L$ as part of an infinite system.
Now, the standard thermodynamic relation $\langle\delta
\hat{N}^2\rangle=k_BT\,\partial N(\mu)/\partial\mu$ in this effectively grand
canonical situation seems to indicate that these fluctuations vanish at zero
temperature which would imply a trivial result $\O\equiv 1$ in the Gaussian
approximation.  This is not true, however, because the relation only applies in
the thermodynamic limit and neglects boundary terms. For a careful calculation
of  $\langle\delta \hat{N}^2\rangle$ at zero temperature and in a finite system,
we generalize the analysis of Giorgini et al.~(\cite{gior1998}, see
also~\cite{astr2007,klaw2011}) for Bose gases in a $d=3$ continuum to arbitrary
dimensions. The particle number fluctuations
\begin{equation}
  \langle \delta \hat{N}^2\rangle=S_d\int_0^{2L}\d r\,
  r^{d-1}\tau(r)\bar{n}\nu(r)\ ,
  \label{eq:giorgini}
\end{equation}
in a spherical domain of radius $L$ can then be calculated from
the pair distribution function
\begin{equation}
  \nu(\vec{r})=\delta(\vec{r})+n\left( g^{(2)}(\vec{r})-1\right)=\int\frac{\d^dq}{(2\pi)^2}e^{i\vec{q}\cdot\vec{r}}S(\vec{q})
  \label{eq:structurefactor}
\end{equation}
and the volume $\tau(r)$ of the intersection between two $d$-dimensional balls of radius $L$
separated by the distance $r$. Here, $S_d$ is the surface of a unit sphere in
$d$ dimensions while $S(\vec{q})$ is the standard static structure factor.  

In the superfluid, the zero temperature static structure factor has the
non-analytic behavior $S(\vec{q})\simeq \alpha|\vec{q}|$ at small wave numbers,
characteristic for any compressible phase. Since collective excitations exhaust
the $f$\,sum rule for long wavelengths, the prefactor $\alpha=\hbar/2m\cs$ is,
moreover, completely fixed by the exact sound velocity $\cs$.  As a result of
the non-analytic behavior of $S(\vec{q})$, the associated pair distribution
function exhibits an algebraic decay $\nu(r)\simeq -\alpha/\pi r^{2}$ in $d=1$
and $\nu(r)\simeq -\alpha/2\pi r^3$ in $d=2$ at long distances.  The behavior of
the number fluctuations for  large $L$ is then found to be 
\begin{equation}
 \langle \delta
 N^2\rangle\sim \alpha\bar{n} L^{d-1}\ln(L/\xih)\quad (\text{SF})  \ ,
  \label{eq:dNSF}
\end{equation}
where the effective healing
length $\xih=\hbar/m\cs$ serves as a short-distance cutoff. Within the Gaussian
approximation, therefore, the parity order
\begin{equation}
  \O\sim
  \begin{cases}
    L^{-\pi\alpha\bar{n}} & (d=1)
    \\
    L^{-\pi^2 \alpha \bar{n}L} &(d=2)
  \end{cases}\quad (\text{SF})
  \label{eq:qualiasymp}
\end{equation}
decays to
zero with a power law in the superfluid phase in one dimension while in  two
dimensions, the decay is super-exponential.

In the MI phase, the quadratic number fluctuations are much
smaller and the scaling of $\langle \delta N^2\rangle$ with system size $L$
differs in a qualitative manner from that in the SF.  Indeed, at zero
temperature, incompressibility of the MI implies that the structure factor
vanishes analytically for $q\rightarrow 0$ and thus $S(\vec{q})=\gamma
(q\xi)^2+\cdots$ to leading order in an expansion around $\vec{q}=0$.  Here,
$\xi$ is the characteristic length of the exponential decay of the one-particle
density matrix mentioned in the previous section, while $\gamma$
is a numerical constant which relates the scales appearing in the density
correlation and in the one-particle density matrix.  From perturbation theory,
one finds that in one dimension
$\gamma$ is proportional to $(J/U)^2$ to leading order~\cite{ejim2011}. 

Since the static structure factor is analytic around $q=0$, the pair correlation
function $\nu(r)$ decays exponentially on the characteristic scale $\xi$,
effectively cutting off the integral in Eq.~\eqref{eq:giorgini}.  As a result,
one finds that 
\begin{equation}
  \langle \delta N^2\rangle\to b L^{d-1}
  \quad (\text{MI})
  \label{eq:qualiasympMI}
\end{equation}
 scales like the area of the boundary of the domain with a coefficient
 $b=b_0\gamma\bar{n}\xi$, where $b_0$ is a numerical constant depending on the
 geometry and the precise functional form of $\nu(r/\xi)$.

The results obtained within this Gaussian approximation are---of course---not
exact. Yet, as will be shown in sections~\ref{sec:dual1d} and \ref{sec:dual2d}
below, they give the correct qualitative behavior of the parity order parameter
as a function of the size $L$ of the domain.  On a basic level, therefore, the
different behavior of the non-local parity order in the MI and the SF is a
simple consequence of the fundamental difference between number fluctuations in
an incompressible versus a compressible system~\cite{rach2012,song2012}. In this
context, it is also instructive to note that the underlying scaling
$\langle\delta N^2\rangle\sim \alpha\bar{n} L^{d-1}\ln(L/\xi)$ of the number
fluctuations in the compressible and gapless superfluid compared to $\langle
\delta N^2\rangle\sim L^{d-1}$ in the incompressible and gapped MI are
reminiscent of similar results obtained for the scaling of the entanglement
entropy $S(L)$. The fact that the fluctations of conserved quantities are
closely related to the latter, exhibiting a simple area law for gapped phases,
has been noted by Swingle and Senthil~\cite{swin2011}.  There are, however, a
number of important differences: the entanglement entropy typically scales like
$S(L)=b\, L^{d-1}$ with a non-universal prefactor $b$ even in gapless phases
while the corresponding number fluctuations $\langle \delta N^2\rangle$ have an
additional logarithmic enhancement factor for any phase with a finite
compressibility.  For the entanglement entropy, violations of the area law by a
logarithmic factor of the form $S(L)\sim L^{d-1}\ln(L)$ appear, e.g., in free
fermions with a Fermi surface and also for Landau Fermi liquids, not in a
gapless superfluid, however.  Indeed, as noted by Metlitski and
Grover~\cite{metl2011}, its entanglement entropy $S(L)= b\, L^{d-1}+\Delta S$
exhibits an {\it additive}\/---not multiplicative---logarithmic contribution
$\Delta S=\ln{\left(\rhos L^{d-1}/\cs\right)}/2$ which is universal, i.e., it
only depends on the superfluid stiffness $\rhos$ and the sound velocity $\cs$ as
effective low energy constants.
    
The simplicity of the Gaussian approximation for studying the parity order also
allows for a straightforward discussion of how the above results are affected by
a finite temperature.  For a neutral SF, the structure factor at temperatures
$k_BT\ll m\cs^2$ reads~\cite{land1980} $S(\vec{q})\simeq
\alpha|\vec{q}|\coth(\cs|\vec{q}|/2\kB T)$. The characteristic length scale at
which the zero temperature result $S(\vec{q})\simeq \alpha|\vec{q}|$ is equal to
the $q=0$ thermal value $S(\vec{q}=0)=2\kB T\alpha/\cs$ is therefore given by
$r_T=\hbar \cs/2\pi \kB T=\lambda_T^2/\xih$, where $\lambda_T$ is the thermal
wavelength and $\xih=\hbar/m\cs$ the effective healing length.  Since the parity
order effectively probes number fluctuations at a finite wave vector $q\simeq
2\pi/L$ the zero temperature results remain valid as long as $r_T\gg L$. Since
typical values of the healing length are of order $\xih\sim 1\,\upmu$m, the
characteristic scale of $r_T$ becomes larger than a lattice spacing at
temperatures below $T\approx 70\,$nK. 

An equivalent reasoning can be applied in the Mott insulating phase, where the
thermal behavior of the structure factor $S(\vec{q}=0)=\bar{n}\kB T\kappa_T$
involves the compressiblity $\kappa_T$. For dimensional reasons, the latter must
be of the form $\kappa_T=(\bar{n}\Delta)^{-1}f(\beta\Delta)$ with the Mott gap
$\Delta$ and a function $f(\beta\Delta)\sim e^{-\beta\Delta}$ which has a
thermally activated form.  The zero temperature results are then valid as long
as the domain size $L$ satisfies 
\begin{equation}
  L\lesssim\, \lambda_T\xi \sqrt{\frac{m\Delta}{f(\beta\Delta)}} \, .
  \label{eq:MIcrossover}
\end{equation}
In the interesting regime $T\ll\Delta$ this is easily satisfied due to the
exponential form of $f(\beta\Delta)$. The actual constraint is rather the
condition $\beta\Delta\gg 1$ which becomes increasingly hard to satisfy as one
approaches the critical point.

\section{Perturbative analysis in the MI}
\label{sec:pert}

Deep in the MI, where $J/U\ll 1$, exact results for the parity order can be obtained by 
a systematic perturbation theory around the atomic limit $J=0$ of the Bose-Hubbard model. 
Here we make use of a method developed by van Dongen~\cite{vand1994}, 
which is an extension of a formalism due to Harris and Lange~\cite{harr1967}.
The basic idea is to construct a canonical transformation of the creation and
annihilation operators, $\hat{a}^\dagger=e^{\hat{S}}\hat{b}^\dagger
e^{-\hat{S}}$ with an anti-hermitian generator $\hat{S}$ such that in the new
basis, the occupation numbers and hence the interaction part remain
invariant under hopping.  Explicitly, one writes the Hamiltonian in the form
$\hat{H}=U\hat{D}+\hat{K}$ and defines the hopping term for the new particles
\begin{equation}
  \hat{T}=-J\sum_{\langle i,j\rangle}\hat{b}_i^\dagger\hat{b}_j
  \quad\Leftrightarrow\quad
  \hat{K}=e^{\hat{S}}\hat{T}e^{-\hat{S}}\ .
  \label{eq:KfromS}
\end{equation}
The requirement that $\hat{D}$ is invariant under hopping is obeyed provided it is
a constant of motion, i.e., $[\hat{H},\hat{D}]=0$. This condition fixes the transformation
$\hat{S}$, which may be calculated order by order in a systematic expansion
\begin{equation}
  \hat{S}=\sum_{n\geq 1}\frac{1}{U^n}\hat{S}_n
   ,
  \label{eq:expandST}
\end{equation}
in powers of $1/U$.
By expanding the exponential and substituting the expansion of the operator
$\hat{S}$, the POP takes the form
\begin{multline}
  \O =   1 +   e^{-i \pi N_0}
  \bigg\{  \sum_{k} \frac{\left(i \pi\right)^k}{k!}
  \bra{\Phi_0}  \left[ \frac{1}{U} \hat{S}_1 +\frac{1}{U^2} \hat{S}_2+\cdots
  ,\left(\hat{N}\left(L\right) \right)^k\right]  \ket{\Phi_0} 
  \\
  +\sum_{k} \frac{\left(i \pi\right)^k}{k!} \bra{\Phi_0}
  \frac{1}{2!} \left[\frac{1}{U} \hat{S}_1,\left[\frac{1}{U}
  \hat{S}_1,\left(\hat{N}\left(L\right)\right)^k\right] \right]\ket{\Phi_0}
  +\cdots  \bigg\}\ ,
  \label{eq:Opert}
\end{multline}
where $\hat{N}(L)$ is the total number operator within the domain \D and $N_0$
is the eigenvalue of $\hat{N}(L)$ in the atomic limit ground state
$\ket{\Phi_0}$. Since $|\Phi_0\rangle$ is an eigenstate of $\hat{N}(L)$, the
first term in the curly bracket vanishes at all orders. As a consequence, to
calculate the POP at order $n$ in $J/U$, one needs to determine the operators
$\hat{S}_1,\dots,\hat{S}_{n-1}$. 

Specifically, substituting $\hat{S}_{1}$ from equation~\eqref{eq:S1}
(cf.~\ref{app:pert}) and evaluating the commutators, one finds
the result~\eqref{eq:Oleadingorder} stated already in section~\ref{sec:quali}.
In one dimension, we have continued the perturbative expansion up to fourth
order in $J/U$. For reasons of space, the details of this calculation are
deferred to~\ref{app:pert} and we only give the result here:
\begin{equation}
  \O=1-8\bar{n}(\bar{n}+1)\left( \frac{J}{U} \right)^2
  -\frac{4}{9}\bar{n}(\bar{n}+1)[\bar{n}(473\bar{n}+217)-234]\left(
  \frac{J}{U}
  \right)^4+\cdots\ .
  \label{eq:Oorderfour}
\end{equation}
Just as the leading order term, the next-to-leading correction is independent of
the domain size $L$ as long as the latter is larger than the order of
perturbation. Hence, $\O$ is independent of $L$ for sufficiently small $J/U$,
in agreement with a DMRG
calculation in~\cite{endr2011} where $\O$ was found to be essentially
independent of $L$ for $J/U\lesssim 0.1$ for domain sizes ranging from 1 to 60.

\section{Duality transformation}
\label{sec:dual}

In the following, we show that within a reduced description of the SF to MI
transition at fixed density in terms of a quantum rotor model~\cite{sach1999},
exact results for the parity order parameter can be obtained from analyzing
a $(d+1)$-dimensional classical lattice model. This
is particularly interesting in the $d=2$ case where the parity order \O can  be
expressed in terms of an equal time Wilson-loop in a nontrivial $U(1)$ gauge
field which is dual to the original lattice boson model. Our aim is twofold:
first of all, unlike the qualitative arguments presented in
section~\ref{sec:quali}, the duality transformation permits us to derive exact
results for the large $L$ behavior of \O in both the MI and SF phases within one
unified framework. From a different point of view, however, the duality of
the lattice boson model to a dynamical gauge field in $2+1$ dimensions can also
be read the other way, where ultracold atoms in an optical lattice provide a
quantum simulator of a nontrivial lattice gauge theory. 

The starting point for the mapping is based on the realization that the SF to MI
transition at fixed density is driven by phase fluctuations
only~\cite{fish1989}. To isolate these from the full Bose-Hubbard Hamiltonian,
it is convenient to consider the limit of large filling $\bar{n}\gg 1$, where
the boson operators may be rewritten in a density-phase representation,
$\hat{a}_j\simeq
e^{i\hat{\phi}_j}\sqrt{\bar{n}+\delta\hat{n}_j}$~\cite{fish1989}.  In the limit
$\bar{n}\gg 1$, the number fluctuations $\delta\hat{n}_j$ can be expanded up to
second order and the Bose-Hubbard Hamiltonian is thus transformed into the
Hamiltonian
\begin{equation}
  H_{\text{J}}=\frac{U}{2}\sum_{\vec{x}}\hat{n}_{\vec{x}}^2+\EJ
  \sum_{\vec{x},\vec{u}}[1-\cos(\hat{\phi}_{\vec{x}+\vec{u}}-\hat{\phi}_{\vec{x}})]
  \label{eq:HJ}
\end{equation}
of a system of quantum rotors at each lattice site $\vec{x}$ with a discrete 
eigenvalue spectrum $0,\pm 1,\pm 2,\ldots$ which are coupled to nearest neighbors 
$\vec{x}+\vec{u}$ by a Josephson energy $\EJ=2\bar{n}J$.   
(Note that we have redefined $\delta\hat{n}_{\vec{x}}\to\hat{n}_{\vec{x}}$, 
i.e.,  $\hat{n}_{\vec{x}}$
is now the deviation from the average local occupation number).  

Following a standard procedure, the partition function associated with the
quantum rotor model~\eqref{eq:HJ} can now be expressed in terms of
a path integral by dividing the interval $[0,\beta]$ into $N_{\tau}$ subintervals of width
$\varepsilon=\beta/N_{\tau}$. Inserting a complete set of number states at each
step, one thus obtains a discretized path integral representation of the form
\begin{equation}
  Z_{\text{J}}=\sum_{n_{\vec{x},j}\in\mathbb{Z}}\, 
  \langle \{ n_{\vec{x},1}\} \vert e^{-\varepsilon\hat{H}_{\text{J}}}\vert \{
  n_{\vec{x},2}\} \rangle \cdots
  \langle \{ n_{\vec{x},N_{\tau}}\} 
  \vert e^{-\varepsilon\hat{H}_{\text{J}}}\vert
  \{ n_{\vec{x},1}\} \rangle\, ,
      \label{eq:ZJ1}
\end{equation}
where the notion $\{ n_{\vec{x},j}\}$ is a reminder of the fact that---at given
$j=1,\ldots, N_{\tau}$---there are $N^d$ variables $n_{\vec{x},j}\in\mathbb{Z}$
for $\vec{x}\in (1,\ldots, N)^d$ that have to be summed over all integers
$\mathbb{Z}$.  In the limit $\varepsilon\to 0$, the non-commuting terms in
$\exp(-\varepsilon\hat{H}_{\text{J}})$ can be factorized to give 
\begin{equation}
  Z_{\text{J}}=\sum_{n_{\vec{x},j}\in\mathbb{Z}}\, 
  \prod_{\vec{x},j,\vec{u}}\, \left( e^{-\varepsilon
  U n_{\vec{x},j}^2/2}
  \langle \{ n_{\vec{x},j}\} \vert e^{-\varepsilon
  \EJ[1-\cos(\hat{\phi}_{\vec{x}+\vec{u}}-\hat{\phi}_{\vec{x}})]} 
  \vert \{ n_{\vec{x},j+1}\}
  \rangle\right)\, .
  \label{eq:ZJ2}
\end{equation}
The matrix elements of the Josephson coupling terms can now be simplified by
using the so called Villain approximation\footnote{a constant prefactor
in~\eqref{eq:Villain} is suppressed because it only gives an irrelevant overall
shift of the free energy.}~\cite{vill1976}  
\begin{equation}
  \exp\{-\varepsilon
  \EJ[1-\cos(\hat{\phi}_{\vec{x}+\vec{u}}-\hat{\phi}_{\vec{x}})]\}
  \simeq
  \sum_{m_{\vec{x},\vec{u}}}
  \exp\left(-\frac{m_{\vec{x},\vec{u}}^2}{2\varepsilon
  \EJ}-im_{\vec{x},\vec{u}}(\hat{\phi}_{\vec{x}+\vec{u}}-\hat{\phi}_{\vec{x}})\right)\, .
  \label{eq:Villain}
\end{equation}
Since the fundamental symmetry $\phi\to\phi+2\pi$ due to the discreteness of the
boson number is retained, this leaves the physics qualitatively unchanged. In
one spatial dimension, this approximation introduces one integer $m_{lj}$
per lattice site, in two spatial dimensions, it introduces two integers
$\vec{m}_{\vec{x},j}=(m_{\vec{x},x,j},m_{\vec{x},y,j})$ at each lattice site.
 
For each given $j$, one now uses the fact that
\begin{equation}
  \prod_{\vec{x}}\, \exp\left(-im_{\vec{x},\vec{u},j}
  (\hat{\phi}_{\vec{x}+\vec{u}}-\hat{\phi}_{\vec{u}})\right)
  =\prod_{\vec{x}}\,
  \exp\left(i(m_{\vec{x},\vec{u},j}-m_{\vec{x}-\vec{u},\vec{u},j})\,\hat{\phi}_{\vec{x}}\right)
  \label{eq:Villain2}
\end{equation}
for a periodic chain with
$\hat{\phi}_{\vec{x}+N\vec{u}}=\hat{\phi}_{\vec{x}}$ and the identity
$\langle n'|e^{im\hat{\phi}}|n\rangle=\delta_{n',n+m}$ since the operator
$e^{im\hat{\phi}}$ shifts the particle number by $m$. The resulting partition
function
\begin{equation}
  Z_{\text{J}}=\sum_{\substack{n_{\vec{x},j}\in\mathbb{Z} \\
  m_{\vec{x},\vec{u},j}\in\mathbb{Z}}}\,
  \exp{\left[-\frac{\varepsilon U}{2}\sum_{\vec{x},j} n_{\vec{x},j}^2
  -\frac{1}{2\varepsilon \EJ} 
  \sum_{\vec{x},\vec{u},j} m_{\vec{x},\vec{u},j}^2\right]}
  \prod_{\vec{x},\vec{u},j}\,
  \delta_{\nabla_{\vec{x}}\cdot \vec{m}_{\vec{x},j},-\nabla_{\tau}
  n_{\vec{x},j}} 
  \label{eq:ZJ3}
\end{equation}
has a Gaussian form, however the variables $n_{\vec{x},j}$ and
$\vec{m}_{\vec{x},j}$  are integer-valued and are connected by the constraint
$\nabla_{\vec{x}}\cdot \vec{m}+\nabla_{\tau} n=0$, where $\nabla_{\vec{x},\tau}$
denotes the discrete derivative on the dual lattice of links along the physical
directions $\vec{x}$ and the 'time' direction $\tau$. Thus, the variables
$n_{\vec{x},j}$ and $\vec{m}_{\vec{x},j}$ together form a divergenceless
$(d+1)$-dimensional integer vector field $\vec{n}\equiv (n,\vec{m})$. In $d=1$,
this constraint may be resolved by introducing a single integer field
$h_{x,\tau}$ such that $n=\nabla_x h$ and $m=-\nabla_\tau h$. The partition
function then becomes that of the discrete Gaussian model with height variable
$h$ which describes the roughening transition of a 2d interface.
Moreover, the POP is mapped on a two-point correlation function of the local
variable $\mathcal{O}(x)=\exp(i\pi h(x))$. This model is discussed
in section~\ref{sec:dual1d}.

In $d=2$, the constraint is automatically satisfied if one introduces a three-component 
vector potential $\vec{a}$ such that $\vec{n}=\nabla\wedge\vec{a}$, where the lattice
curl is defined as
$(\nabla\wedge\vec{a}_{\vec{x}})_i=\sum_{j,k}\varepsilon_{ijk}(a_{\vec{x}-\vec{\hat{k}},k}-a_{\vec{x}-\vec{\hat{\jmath}}-\vec{\hat{k}},k})$.
The partition function then becomes that of a $(2+1)$-dimensional $U(1)$ gauge
theory, as will be discussed in detail in section~\ref{sec:dual2d}.
Remarkably, under the duality transformation the parity order parameter is
mapped onto an equal time Wilson loop~\cite{wils1974}
\begin{equation}
  \O=\left\langle \exp\left[
  i\pi\sum_{\vec{x}\in\D}\d^2x\,(\nabla\wedge\vec{a})_\tau
  \right]\right\rangle
=\left\langle
\exp\left[i\pi\sum_{\vec{x}\in\partial\D}(\Delta\vec{x})\cdot\vec{a}
\right]\right\rangle
\ ,
  \label{eq:wilson}
\end{equation}
where we have used the discrete version of Stokes's theorem on the last line,
$\partial\D$ is the boundary of the domain \D and $\Delta\vec{x}$ is a unit
vector directed along the boundary in the positive mathematical sense.  The
Wilson loop is a gauge-invariant quantity which is often used to characterize
phases in gauge theories~\cite{kogu1979}.  As will be shown in
section~\ref{sec:dual2d}, in the present case, the transition between a phase
with massless and one with massive 'photons' in the underlying $U(1)$-gauge
theory predicts qualitatively different behavior of the parity order in the SF
and MI phases of the original lattice Boson model, consistent with the
qualitative considerations discussed in section~\ref{sec:quali}.  

\section{Discrete Gaussian interface model}
\label{sec:dual1d}

As shown in the previous section, in the $d=1$ case the partition function
can be represented in terms of a single integer $h$ on each
lattice site such that $(n,m)=(\nabla_x h,-\nabla_\tau h)$. 
Choosing $\varepsilon=1/U$,\footnote{Note that
the Trotter decomposition in Eq.~\eqref{eq:ZJ1} is usually performed for finite
$\beta$ and thus $N_\tau=\beta/\varepsilon\rightarrow\infty$ requires
$\varepsilon\rightarrow 0$. Here we keep $\varepsilon$ finite but consider the
limit of zero temperature $\beta\rightarrow\infty$.}
the resulting partition function 
\begin{equation}
    Z_{\text{DG}}=\sum_{\{h_{lj}\}}\exp\left[ 
      -\frac{1}{2}
        \sum_{l,j}\left\{ (\nabla_x h_{lj})^2
	  +\frac{U}{\EJ}(\nabla_\tau h_{lj})^2 \right\}\right]
	   \label{eq:ZDG}
\end{equation}
defines an anisotropic discrete Gaussian (DG) model for an integer valued
height variable $h_{lj}$ above a two-dimensional, perfectly flat 
interface $h_{lj}\equiv 0$ (note that an overall shift $h_{lj}\rightarrow h_{lj}+\mathbb{Z}$ 
of this reference plane is irrelevant). This mapping has been used 
earlier in the context of the SF-MI transition in one dimension by one of the present authors~\cite{zwer1989}. 

The DG model is a classical model which exhibits a  phase transition from a
smooth interface in the regime where $U$ dominates to a rough phase in the limit
$\EJ\gg U$.  The smooth phase, which corresponds to the MI in the original
quantum rotor model, is characterized by a finite dimensionless step free energy
$f_{\text{s}}$~\cite{fish1983} which is related to the Mott gap $\Delta\mu$ of
the dual model~\eqref{eq:HJ} by $2f_{\text{s}}=\Delta\mu/U$~\cite{zwer1989}.
The dimensionless step free energy is a decreasing function of $\EJ/U$ and
reaches $f_{\text{s}}\equiv 1/2$ at $\EJ=0$. In this limit an additional boson
is described by a step of unit height which is parallel to the $\tau$~axis,
i.e., the boson world lines exhibit no quantum fluctuations.

When $\EJ/U\sim 1$, the model is essentially isotropic. To render this manifest,
it is convenient to choose $\varepsilon=1/\sqrt{\EJ U}$ so that
\begin{equation}
    Z_{\text{DG}}=\sum_{\{h_{lj}\}}\exp\left[ 
    -\frac{1}{2}\sqrt{\frac{U}{\EJ}}
        \sum_{l,j}\left\{ (\nabla_x h_{lj})^2
	  +(\nabla_\tau h_{lj})^2 \right\}\right]\ .
	   \label{eq:ZDGiso}
\end{equation}
In this representation, the ratio $\TDG=\sqrt{\EJ/U}$ of kinetic and interaction
energy in the underlying quantum rotor Hamiltonian~\eqref{eq:HJ} plays the
role of an effective temperature.  The discrete Gaussian model~\eqref{eq:ZDGiso} 
is known to have a roughening transition of the Kosterlitz-Thouless type~\cite{chui1976} 
at a critical temperature $\TR\simeq 0.73$.  In the smooth phase, the mean
square surface displacement 
\begin{equation}
\Delta h^2(L)=\langle(h_{lj}-h_{l'j'})^2\rangle
  \label{eq:hsquare}
\end{equation}
remains finite as the distance $L=|(l,j)-(l',j')|$ between two points on the
surface approaches infinity. By contrast, the rough phase of the discrete
Gaussian model at $\tilde{T}_{\text{DG}}>\TR$ is characterized by a
logarithmically divergent $\Delta h^2(L)\sim\ln(L)$. 

These results on the discrete Gaussian model, for which the existence of a
Kosterlitz-Thouless transition has been proven rigorously by Fr\"ohlich and
Spencer~\cite{froh1981}, can now be translated back to understand the nature of
non-local order in the original Bose-Hubbard or quantum rotor model.  In
particular, the qualitative results that were derived in section~\ref{sec:quali}
within a  Gaussian approximation from considering the number fluctuations within
a domain of linear size $L$ can now be put on a rigorous footing. This relies on
the fact that number fluctuations in the original model of bosons hopping on a
lattice are transformed, via the duality, to fluctuations of the normal vector
of the 2d interface by $n=\nabla_x h$. As a result, the nonlocal order parameter
defined in Eq.~\eqref{eq:OP} translates into the characteristic function
\begin{equation}
  \langle\mathcal{O}^2(L)\rangle=\langle
\exp(i\pi[h(L)-h(0)])\rangle
\label{eq:maptwopt}
\end{equation}
of the probability distribution
$p(h,|(l,j)-(l',j')|)=\langle\delta(h_{lj}-h_{l'j'}-h)\rangle$ that the height
variables at two sites at a distance $L$ differ by $h$.  

For a detailed understanding of the behavior of the parity order parameter, we
can now use exact results on the classical roughening transition of
two-dimensional interfaces. Specifically, within the smooth phase,
Forrester~\cite{forr1986} has obtained the exact probability distribution for
the height of a lattice site near the center in the body-centered solid-on solid
(BCSOS) model for fixed boundary conditions and the thermodynamic limit: it
turns out to be a discrete Gaussian distribution
$p(h)=\mathcal{N}e^{-(h-1/2)^2/2\sigma^2}$, where $\mathcal{N}$ is a
normalization constant, and with $\sigma^2=C/\sqrt{1-T/T_{\text{R}}}$ where
$C=\sqrt{2/\ln 2}$. The non-zero expectation of this distribution stems from the
fact that in the BCSOS model, one has to deal with two sublattices where the
sites of one take only even values while the sites of the other take only odd
values. Since the outermost sites are fixed at values $0$ and $1$ (according to
the sublattice), the height in the center may equivalently be seen as the height
\emph{difference} to the border, or, in the thermodynamic limit, as the
infinite-distance limit of this difference. Hence, the characteristic function
of this probability distribution is just the two-point correlation function
discussed above within the DG model.  Since the DG and the BCSOS model are in
the same universality class, we may conclude that in the limit of infinite
distance, the probability distribution for the height difference between two
points in the DG model equally becomes a discrete Gaussian distribution
$p(\Delta h=n)=\mathcal{N} e^{-n^2/2\sigma^2}$ with\footnote{Note that a priori,
$C_{\text{DG}}\neq C$~\cite{forr1986}. The numerical values of the roughening
temperatures are model-specific as well.}
$\sigma^2=C_{\text{DG}}/\sqrt{1-\TDG/\TR}$. Using the Poisson formula, one then
obtains the characteristic function 
\begin{equation}
  \sum_{n\in\mathbb{Z}}p(n)e^{ikn}
  =\frac{\sqrt{2\pi\sigma^2}}{\sum_{n\in\mathbb{Z}}e^{-n^2/2\sigma^2}}
  \sum_{m\in\mathbb{Z}}e^{-\sigma^2(k-2\pi m)^2/2}\ ,
  \label{eq:charfctgauss}
\end{equation}
i.e., the characteristic function is a periodic sum of Gaussians centered on
$0,\pm 2\pi, \pm 4\pi, \dots$. The prefactor is just the ratio of the
normalization constants of the continuous and the discrete Gaussian distribution
and tends to unity for $\sigma^2\gg 1$. As $\TDG$ approaches $\TR$ from below,
$\sigma^2$ diverges and the individual peaks of the
characteristic function become very narrow. There are then only two
contributions to the POP (from the peaks centered on $0$ and $2\pi$) which goes
to zero as
\begin{equation}
  \langle \mathcal{O}^2(\infty)\rangle
  =p(k=\pi)\sim 2\exp\left( -\frac{\pi^2 C_{\text{DG}}}{2\sqrt{1-\TDG/\TR}} 
  \right)\ ,
  \label{eq:BKTscaling}
\end{equation}
confirming the conjecture $\O\sim\exp\{-A[(J/U)_{\text{c}}-(J/U)]^{-1/2}\}$ made
in~\cite{endr2011} on how the POP reaches zero as one approaches the critical
point from the smooth phase (close to the critical point, writing $\O$ as a
function of $J/U$ rather than $\TDG\propto \sqrt{J/U}$ only affects the
non-universal constant $A$). 

To understand the behavior of the parity order within the superfluid, which
corresponds to the rough phase of the associated 2d classical interface model,
it is convenient to use  the equivalence between the DG model and the classical
two-dimensional Coulomb gas (CG) derived by Chui and Weeks~\cite{chui1976}.
Within the 2d CG picture,  the underlying SF-MI transition is translated into a
phase transition of the Kosterlitz-Thouless type between an insulating phase
where charges are bound---the SF phase of the original model---and a metallic
phase of effectively free charges which describes the MI.  The metallic phase is
characterized by a divergent polarizability $\int d^2r\, r^2p(r)$, where $p(r)$
is the probability distribution function for the distance between a pair of
opposite unit charges added to the system.  Indeed, as shown by Chui and
Weeks~\cite{chui1976}, for $0<\xi<2\pi$ the two-point correlation function of
the DG model maps on the partition function of the Coulomb gas in the presence
of two opposite charges $\pm\xi$ at a distance $r=|\vec{r}_1-\vec{r}_2|$,
\begin{equation}
  \left\langle e^{i\xi[h(\vec{r}_1)-h(\vec{r}_2)]}\right\rangle
  =e^{-\beta_{\text{CG}}F(r,\xi)}=p(r)\ ,
  \label{eq:freeenergy}
\end{equation}
where $F(r,\xi)$ is the free energy of the neutral Coulomb gas in the presence
of the added charges. It is well known~\cite{minn1981} that when these added
charges have unit strength [within the mapping by Chui and Weeks, a unit charge
corresponds to $\xi=2\pi$ in $F(r,\xi)$], 
\begin{equation}
  e^{-\beta_{\text{CG}}F(r,2\pi)}
  =\exp\left( -\frac{e^2}{\varepsilon_0}\ln r \right)=r^{-e^2/\varepsilon_0}\ ,
  \label{eq:screening}
\end{equation}
with the dielectric constant $\varepsilon_0$ which characterizes the insulating
phase of the Coulomb gas. At the unbinding transition, the polarizability
diverges, i.e., the critical value of the dielectric constant obeys
$(e^2/\varepsilon_0)_{\text{c}}=4$. In our case of $\xi=\pi$, we are dealing
with a pair of half-unit charges\footnote{To avoid possible confusion, we stress
that the right-hand side of Eq.~\eqref{eq:freeenergy} is well-defined for
arbitrary $\xi$, but the equality to the left-hand side (i.e., the existence of
the mapping) is restricted to values $0<\xi<2\pi$, and thus to fractional
charges. For $\xi = 2\pi n$ with integer $n$, one trivially has
$e^{i\xi[h(\vec{r}_1)-h(\vec{r}_2)]}=1$.}. As a result the POP 
\begin{equation}
\O\sim r^{-e^2/4\varepsilon_0}\ ,   
  \label{eq:POPscalingSF}
\end{equation}
decays algebraically in the SF phase of the original lattice Bose model, and the
mapping to the Coulomb gas gives us the exact value of the universal jump of the
exponent at the roughening temperature: for $\TDG\rightarrow\TR$ from above, the
exponent $e^2/4\varepsilon_0$ approaches one before jumping to zero, a
consequence of the universal jump of the superfluid stiffness at the SF-MI
transition of the 1d quantum rotor model (for a more detailed discussion
see~\cite{zwer1990}). 

\section{$\boldsymbol{U(1)}$ gauge theory in $\boldsymbol{2+1}$ dimensions}
\label{sec:dual2d}

In $d=2$, the duality mapping discussed in section~\ref{sec:dual} and the
introduction of the three-component vector potential $\vec{a}$ with
$\vec{n}=\nabla\wedge\vec{a}$ lead to a $(2+1)$-dimensional $U(1)$ gauge
theory~\cite{wen2004,pesk1978,fish1989a,dasg1981,herb1998} (for an alternative
approach to this type of duality mapping using the operator formalism,
cf.~\cite{bale2005}). Similarly to the 1d case, it is convenient to choose
$\varepsilon=1/\sqrt{U\EJ}$, which results in a partition function of the form
\begin{equation}
 Z=\sum_{\{\vec{a}\}}\exp\left(
  -\frac{1}{2}\sqrt{\frac{U}{\EJ}} \sum_{\vec{x},\tau}(\nabla\wedge\vec{a})^2
  \right)\ .  
  \label{eq:ZJfinal} 
  \end{equation}
Since $(\nabla\wedge\vec{a})^2=\vec{e}^2+b^2$ is the energy density associated
with a two-component 'electric field' $e_i=\nabla_\tau a_i-\nabla_i a_\tau$ and
a scalar 'magnetic field' $b=\nabla_x a_y-\nabla_y a_x$, this partition function
appears to describe the free field theory of pure electrodynamics in
$(2+1)$~dimensions~\cite{wen2004}. It is gauge invariant since the
action only depends on the curl of $\vec{a}$ and is thus unchanged if the
lattice gradient of an arbitrary function of $\vec{x}$ and $\tau$ is added to
$\vec{a}$. What makes the model in Eq.~\eqref{eq:ZJfinal} nontrivial is the fact
that all fields take only integer values on a discrete space-time lattice.
The SF to MI transition of the underlying lattice Bose model shows up at the
level of the dual $U(1)$ gauge theory as a transition between 
a phase in which $\vec{a}$ really is a free field and the photon is massless
and one with massive photons. 

In order to understand the physical meaning of this transition in terms of the 
gauge field degrees of freedom, it is convenient to consider the
equal-time one-body density matrix in the quantum rotor model~\eqref{eq:HJ},
which is mapped onto a ratio of two gauge field partition functions
\begin{equation}
  \frac{\langle \hat{a}^\dagger(\vec{x})\hat{a}\rangle}{\bar{n}}
  =\langle
  e^{i\hat{\phi}(\vec{x})}e^{-i\hat{\phi}(0)}\rangle
 = \frac{Z[\vec{x},0]}{Z}
 =\exp(-\Delta F[\vec{x},0])
 \ ,
  \label{eq:corrfctmap}
\end{equation}
where $Z[\vec{x},0]$ differs from $Z$ in that the constraint
$\nabla\cdot\vec{n}(\vec{y})=0$ is replaced by
$\nabla\cdot\vec{n}(\vec{y})=\delta_{\vec{y},\vec{x}}
-\delta_{\vec{y},0}$~\cite{herb2007}.  Physically, this corresponds to a
configuration with a pair of oppositely charged magnetic monopoles situated at $\vec{x}$ and $0$.
The fact that the one-body density matrix approaches a constant in the SF 
and decays exponentially in the MI thus leads to a fundamentally different
behavior of the dimensionless free energy increase 
\begin{equation}
  \Delta F[\vec{x},0]=
  \begin{cases}
    |\vec{x}|/\xi & (\text{MI})
    \\
    \text{const.}-\frac{\cs}{4\pi\rhos|\vec{x}|} & (\text{SF})
  \end{cases}
  \label{eq:DeltaF}
\end{equation}
associated with the introduction of a monopole--antimonopole pair at 
distance $\vec{x}$ in the dual gauge theory (here, \rhos and \cs denote the
superfluid stiffness and sound velocity, respectively). In particular, 
the exponential decay of the one-body density matrix in the MI phase
translates to a linear confinement while in the SF phase the monopoles interact
via a 3d Coulomb potential.  

An effective low energy description of the gauge theory 
which properly accounts for the two different phases is provided by
a Gaussian model of the form
\begin{equation}
  S=
  \frac{1}{2\tilde{T}}
  \int\d^3x\,\left\{ 
  [\nabla\wedge\vec{a}(\vec{x})]^2+\frac{1}{\xi^2}[\vec{a}(\vec{x})]^2
  \right\}\ ,
  \label{eq:SeffSC}
\end{equation}
where $\vec{a}$ is now treated as a \emph{continuous} variable and
$\tilde{T}\simeq\sqrt{ E_{\text{J}} /U}$ is a renormalized dimensionless
temperature or coupling constant, similar to the one in the previous 
section. In the SF regime, where $\tilde{T}$ is above a critical value of  order one,
the gauge field is in its Coulomb phase where $\xi=\infty$. 
Elementary excitations are then massless photons,
which are just the phonons of the superfluid with linear dispersion
$\omega=\cs|\vec{q}|$ (note that 'photons' in $(2+1)$-dimensional
electrodynamics have no polarization degrees of freedom). By contrast, 
in the MI for small values of $\tilde{T}$, the gauge field is in a confining
phase, with $\xi=1/m$ finite. The photons thus acquire a mass $m$ and now
represent the elementary  particle--hole excitations of the MI with dispersion
$\omega =\cs\sqrt{m^2\cs^2+q^2}$ (at the transition, one has $\cs=4.8J$ for
the Bose--Hubbard model with $\bar{n}=1$~\cite{capo2008}).
When evaluating expectations for the massless phase, it is convenient to keep a
finite $\xi$ during the calculation and only take the limit $\xi\rightarrow
\infty$ at the end of the calculation, which avoids the necessity of
introducing an explicit gauge-fixing term~\cite{zee2010}. This can be seen
explicitly in the correlation function of the vector potential, which reads  
\begin{equation}
  \langle a(\vec{q})_j a(\vec{q'})_k\rangle
  =
  \tilde{T}
  (2\pi)^3\delta(\vec{q}+\vec{q'})\left[ 
  \frac{[P_t(\vec{q})]_{jk}}{\vec{q}^2+\xi^{-2}}
  +\frac{[P_l(\vec{q})]_{jk}}{\xi^{-2}}
  \right]\ ,
  \label{eq:SCcorrelator}
\end{equation}
where $[P_t(\vec{q})]_{jk}=\delta_{jk}-q_jq_k/\vec{q}^2$ and
$[P_l(\vec{q})]_{jk}=q_jq_k/\vec{q}^2$ are the components of the transverse and 
longitudinal projector with respect to $\vec{q}$, respectively. Since the 
effective model~\eqref{eq:SeffSC} is Gaussian, the calculation 
of the parity order parameter from the Wilson loop in Eq.~\eqref{eq:wilson} is now 
easy and gives\footnote{The resulting scaling of the parity order will thus turn
out to be the same as found in section~\ref{sec:quali} up to numerical factors,
since both are obtained from a Gaussian model. Note, however, that the mapping
of \O to a Wilson loop in the dual gauge theory provides a proper justification
for these scalings.}
\begin{equation} 
  \O=\exp\left( -\frac{\pi^2}{2}\left\langle
  \left(\int_\D\d^2x\,[\nabla\wedge\vec{a}(\vec{x})]_\tau\right)^2 \right\rangle
  \right)\ . 
  \label{eq:Gausswilson}
\end{equation}
Using \eqref{eq:SCcorrelator}, the expectation appearing in the
exponent in Eq.~\eqref{eq:Gausswilson} reduces to
\begin{equation}
\left\langle
  \left(\int_\D\d^2x\,[\nabla\wedge\vec{a}(\vec{x})]_\tau\right)^2
  \right\rangle
  =
  \tilde{T}
  \int\frac{\d^3q}{(2\pi)^3}\,\frac{q_x^2+q_y^2}{\vec{q}^2+\xi^{-2}}
  \left|\int_\D\d^2x\,e^{i\vec{q}_\perp\cdot\vec{x}}
  \right|^2\ ,
  \label{eq:POPint}
\end{equation}
where $\vec{q}_\perp=(q_x,q_y)$.  Note that $\vec{q}$ has three finite
components whereas $\vec{x}$ is restricted to the $\tau=0$ plane. For a circular
disk of radius $R$, the spatial integral appearing in~\eqref{eq:POPint} gives
\begin{equation} 
  \int_\D\d^2x\,e^{i\vec{q}_\perp\cdot\vec{x}} =
  \frac{2\pi R}{q_\perp}\J_1(q_\perp R) \ .  
  \label{eq:diffraction}
  \end{equation}
 In the deconfined phase, which corresponds to the superfluid, the 
 mass parameter $1/\xi$ is equal to zero and the resulting parity order
 parameter is given by\footnote{The momentum integrals in \eqref{eq:POPint} are
 cut off at $\Lambda\sim\pi$ since the original problem is defined on a lattice
 with lattice constant set equal to unity.}
\begin{equation}
  -\ln\OR\sim \pi^3\tilde{T}R\ln(\pi R)\ .
  \label{eq:POP2dSF}
\end{equation}
in agreement with the qualitative result obtained in Eq.~\eqref{eq:qualiasymp}
within the Gaussian approximation.  Conversely, in the confined phase
corresponding to the MI, The $\vec{q}^2$ is negligible compared with $1/\xi^2$,
leading to a perimeter law
\begin{equation}
  -\ln\OR\sim
  \tilde{T}\xi^2
  2\pi R
  \equiv \frac{R}{\ell_{\mathcal{O}}}\ .
  \label{eq:POP2dMI}
\end{equation}
Thus, the characteristic decay length of the POP scales as
$\ell_{\mathcal{O}}\sim \Delta^2\sqrt{U/J}$ close to the critical point.
Remarkably, a measurement of the parity order, which only involves the
statistics of number fluctuations, therefore allows to extract the Mott gap.

\section{Conclusion \& Outlook}
\label{sec:outlook}

We have presented a detailed study of parity order \O for lattice bosons in both
one and two spatial dimensions, using duality transformations. Consistent with
previous theoretical work~\cite{berg2008} and recent
experiments~\cite{endr2011}, it has been shown that the Mott insulating phase in
one dimension exhibits long range parity order. An intuitive understanding of
this result relies on the observation that for any incompressible phase, the
number fluctuations $\langle\delta \hat{N}^2\rangle\sim L^{d-1}$ in a domain of
size $L$ scale with the area of the boundary $\sim L^{d-1}$.  Using the duality
to a discrete, classical interface roughening problem in two dimensions, these
results have been put on a rigorous footing. In two dimensions, the parity order
again shows qualitatively distinct behavior in the MI and SF phase. In this
case, the variable \O can be expressed in terms of an equal time Wilson loop of
a nontrivial $U(1)$ gauge theory in $2+1$ dimensions.  This is related to the
fact that the density fluctuations in the original lattice Bose model are mapped
to the scalar magnetic field in the dual gauge theory.  A quite interesting
result obtained from this mapping is the fact that the decay of parity order in
the MI allows to measure the Mott gap from the statistics of number
fluctuations. Since experimental measurements of the parity order in two
dimensions are straightforward in principle, this seems a promising route to
infer microscopic information from single site resolution imaging which is very
difficult to obtain otherwise. The detailed numerical factors in the scaling of
\O can unfortunately not be predicted from our effective long wavelength
description of the quantum rotor model. Since we are dealing with a bosonic
system, however, effective numerical methods are available to obtain precise
results in a realistic experimental setup, as has been shown for thermodynamic
properties and excitation energies~\cite{capo2008}.  In particular, numerical
simulations directly deal with the Bose Hubbard model which applies in the
relevant case of low filling $\bar{n}=1$ instead of the qualitatively similar
situation $\bar{n}\gg 1$ that has been studied here within the quantum rotor
model. 

An important open question is, of course, to which extent our results for Bose
Mott insulators can be generalized to the fermionic case, which has also been
realized experimentally with ultracold atoms~\cite{jord2008,schn2008}.  Based on
the qualitative description in terms of the scaling of number fluctuations in
section~\ref{sec:quali}, we expect that the results obtained here carry over to
the fermionic case despite the fact that no duality transformations exist in
this case which allow to connect the parity order with a Wilson loop in a dual
gauge theory. The presence of long range parity order for 1d Mott insulators
also in the fermionic case is consistent  with the results of a recent DMRG
study of the fermionic Hubbard model by Montorsi and Roncaglia~\cite{mont2012}.
A different kind of non-local order characterized by sub-lattice parity was
found in this Model by Kruis et~al.~several years
earlier~\cite{kruis2004,kruis2004a}.

Finally, a quite interesting direction of further research on non-local orders
for cold atoms in optical lattices is connected with the recent realization of a
1d transverse Ising model using a tilted optical lattice~\cite{simo2011}.
Extending this setup to the case of two dimensions, a number of complex phases
may appear depending on the type of lattice and the direction of the
tilt~\cite{piel2011}. A quite intriguing perspective would appear in a setup
that allows to realize a ferromagnetic version of this standard model for
quantum phase transitions~\cite{sach1999}. In the ferromagnetic case, the
transverse Ising model is self-dual in one dimension, while in two dimensions
the dual theory is given by an Ising gauge theory~\cite{kogu1979}. For the
latter, one can define non-local correlation functions as $C(\D)=\langle\prod_{m
\in \D}\hat{S}_m^{x}\rangle$, where $\hat{S}_m^{x}$ is the $x$ component of
the spin operator at site $m$ and $\D$ is an area in two dimensions, in complete
analogy with our definition of the parity order in equation~\eqref{eq:OP}.
After the duality transformation, this observable is transformed into
$C(\D)=\langle\prod_{k \in \partial \D}\hat{\sigma}_k^{z}\rangle$, where
$\hat{\sigma}_k^{z}$ is the $z$ component of the spin operator at site $k$ in
the dual theory and $\partial \D$ are the two sites at both ends of the string
in one dimension or the border of the area in two dimensions. As a result, one
has in one dimension that $\lim_{L\rightarrow \infty}C(L)>0$ for
the paramagnetic phase. In two dimensions, $C(\D)$ is a Wilson-Wegner
loop around a closed path~\cite{wegn1971,kogu1979}, which shows
an exponential scaling with the perimeter of the loop in the
paramagnetic phase and with the enclosed area in the ferromagnetic
phase of the original model~\cite{kogu1979,savi1980}. However, the detection of the
non-local order parameter requires the measurement of the $x$ component
of the spin operator, which in turn requires the single-site resolved
detection of the phase coherence between superposition states with
different on-site occupation numbers, a technique that is so far not
available.

\section*{Acknowledgements}
We acknowledge useful comments by Nigel Cooper and Senthil Todadri. This 
work has been supported by the DFG Forschergruppe 801.

\appendix
\section{Details about the perturbative calculation}
\label{app:pert}

According to the program outlined in Section~\ref{sec:pert}, calculting the POP
to $n$th order in $J/U$ amounts to calculating the operators
$\hat{S}_1,\dots,\hat{S}_{n-1}$. Hence, for the second order result, we only
need to calculate one operator. One finds
\begin{equation}
  \hat{S}_1= {\sum_{D_1,D_2}}' {\frac{1}{D_2-D_1} \hat{P}_{D_2}
    \hat{T}
    \hat{P}_{D_1}}\ ,
  \label{eq:S1}
\end{equation}
where the sums go over the eigenvalues $D$ of the operator $\hat{D}$ and
$\hat{P}_D$ is the projector on the eigenspace corresponding to this eigenvalue.
Moreover, here and in the following, a prime on a sum indicates that values
which make the denominator vanish are excluded from the sum. 

One may imagine the calculation of the expectation \O as summing over all
possible hopping processes where the order in $J/U$ indicates the number of
occurring hops.  For example, at second order, starting from a situation with
uniform filling corresponding to $|\Phi_0\rangle$, all possible processes
consist of a single particle hopping to a neighboring site and back again.
Different contributions stem from the position of the starting lattice site and
its neighbor relative to the domain boundary and from the position of the
operator $\hat{N}(\D)$ in the product of operators.

The next non-vanishing contribution to the POP is of fourth order, so we need to
calculate $\hat{S}_{2,3}$. The former turns out to be given by
\begin{multline}
  \hat{S}_2= {\sum_{D_1,D_2,D_3}}' { \frac{1}{2 \left(D_2-D_1\right)}  \left(
    \frac{1}{D_2-D_3   } - \frac{1}{D_3-D_1   }   \right)
    \hat{P}_{D_2}\hat{T}\hat{P}_{D_3}\hat{T}\hat{P}_{D_1}  } 
    \\
    +{\sum_{D_1,D_2}}' {\frac{1}{\left(D_2-D_1\right)^2} \left( \hat{P}_{D_2}
      \hat{T} \hat{P}_{D_1} \hat{T} \hat{ P}_{D_1} - \hat{P}_{D_2} \hat{T}
      \hat{P}_{D_2} \hat{T} \hat{P}_{D_1}\right)}\ .
    \label{eq:S2}
\end{multline}

\begin{figure}[htbp]
  \centering
  \hfill
    \includegraphics[width=.2\linewidth]{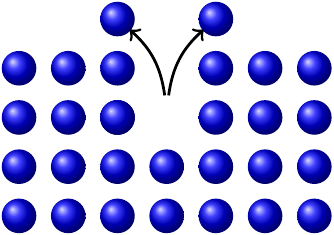}\hfill
    \includegraphics[width=.2\linewidth]{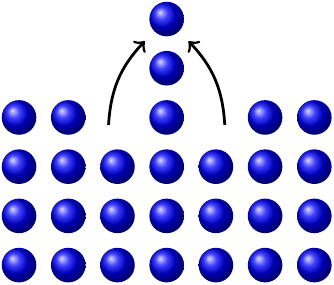}\hfill
    \includegraphics[width=.2\linewidth]{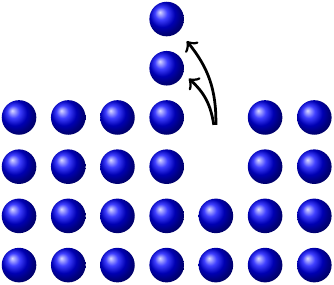}\hfill\phantom{|}
  \caption{(Color online) Possible processes at fourth order where two particles
  start or end on the same lattice site. The images show the configuration after
  two hopping events (the remaining two restore uniform filling) for
  $\bar{n}=4$.}
  \label{fig:pert}
\end{figure}

The expression for $\hat{S}_3$ is too unwieldy to reproduce here, so we limit
ourselves to a description of the involved hopping processes. They can be
grouped into four types:
\begin{itemize}
  \item Back- and forth hoppings of two particles separated by more than two
    lattice sites, i.e., independent second order processes.
  \item Processes where the same particle or hole hops twice before returning to
    its original site.
  \item Processes where a particle hops twice and is then followed by the
    created hole or vice versa.
  \item Processes where two particles start or end on the same lattice
    site before hopping back (cf. Fig.~\ref{fig:pert}).
\end{itemize}
Summing over all contributions finally yields Eq.~\eqref{eq:Oorderfour}.


\end{document}